\documentstyle[12pt,epsfig,amsfonts]{book}
\title{\Huge\bf Radiation Reaction\protect\\ in Classical Field Theory}
\author{Yurij YAREMKO and Volodymyr TRETYAK}
\date{LAP LAMBERT Academic Publishing\\
Heinrich-Bocking-Str. 6-8\\
66121, Saarbr${\rm\ddot u}$cken, Germany
}
\begin{document}

\maketitle

\chapter*{Contents}

\noindent
{\bf Introduction}\hfill 1

\vspace{10pt}

\noindent
{\bf 1. Some basic mathematics}\hfill 11

1.1 Affine space\hfill 12

1.2 Coordinatization of Euclidean space\hfill 15

\hspace{20pt}1.2.1 Cartesian coordinate system\hfill 16

\hspace{20pt}1.2.2 Spherical coordinate system\hfill 17

1.3 Definition of a manifold\hfill 20

\hspace{20pt}1.3.1 Tangent bundle and coordinate bases \hfill 23

1.4 One-forms and tensors\hfill 30

\hspace{20pt}1.4.1 Tensors\hfill 32

1.5 Metric tensor\hfill 33

\hspace{20pt}1.5.1 Raising and lowering of indices\hfill 35

\hspace{20pt}1.5.2 Noncoordinate bases\hfill 36

1.6 $q$-forms, volumes and integration\hfill 38

\hspace{20pt}1.6.1 Hodge star operator\hfill 42

\hspace{20pt}1.6.2 Exterior differentiation\hfill 45

\hspace{20pt}1.6.3 Closed and exact differential forms\hfill 47

\hspace{20pt}1.6.4 Stokes' theorem\hfill 48

\vspace{10pt}

\noindent
{\bf 2. Geometrical aspects of classical electrodynamics}\hfill 53

2.1 Minkowski space\hfill 55

2.2 Poincar\'e group\hfill 63

\hspace{20pt}2.2.1 Group of spacetime translations\hfill 65

\hspace{20pt}2.2.2 Group of spatial rotations\hfill 66

\hspace{20pt}2.2.3 Lorentz group\hfill 70

\hspace{20pt}2.2.4 Poincar\'e group\hfill 74

2.3 {\it Poincar\'e algebra}\hfill 74

2.4 Maxwell's equations\hfill 85

\hspace{20pt}2.2.1 One-form potential and wave equation\hfill 89

\hspace{20pt}2.2.1 {\it Maxwell's equations in integral form}\hfill 90

2.5 The principle of least action\hfill 94

\hspace{20pt}2.5.1 Lagrangian formalism in electrodynamics\hfill 100

2.6 Noether's theorem and conserved quantities\hfill 103

\hspace{20pt}2.6.1 Energy-momentum tensor density\hfill 106

\hspace{20pt}2.6.2 Angular momentum tensor density\hfill 110

2.7 Algebraic properties of the electromagnetic field tensor\hfill 113

\hspace{20pt}2.7.1 Invariants of the electromagnetic field\hfill 113

\hspace{20pt}2.7.2 Eigenvalues and eigenvectors of the field tensor\hfill 
124

\hspace{20pt}2.7.3 Charge in static uniform electromagnetic field\hfill 128

\vspace{10pt}

\noindent
{\bf 3. Green's functions}\hfill 133

3.1 The Dirac delta-function\hfill 134
 
\hspace{20pt}3.1.1 Heaviside stepfunction\hfill 137

\hspace{20pt}3.1.2 Derivative of the delta-function\hfill 138

3.2 Fourier transform\hfill 140

\hspace{20pt}3.2.1 Coulomb potential\hfill 144

3.3 Green's function for d'Alembert operator\hfill 146

\hspace{20pt}3.3.1 Cauchy's residue theorem\hfill 148

3.4 Green's functions in various dimensions\hfill 155

\hspace{20pt}3.4.1 Green's function in three dimensions\hfill 160

\vspace{10pt}

\noindent
{\bf 4. Regularization procedure in classical electrodynamics}\hfill 165

4.1 Radiation corrections to a motion in constant field\hfill 167

4.2 Li\'enard--Wiechert potentials\hfill 170

4.3 Li\'enard--Wiechert fields\hfill 176

\hspace{20pt}4.3.1 Symmetry with respect to time inversion\hfill 180

4.4 Radiation reaction via analysis of fields' asymptotic 
behaviour\hfill 182

\hspace{20pt}4.4.1 Advanced point\hfill 182

\hspace{20pt}4.4.2 Radiative potential and radiation reaction\hfill 184

4.5 Retarded coordinate systems\hfill 188

\hspace{20pt}4.5.1 Dirac's retarded coordinates\hfill 190

\hspace{20pt}4.5.2 Bhabha's retarded coordinates\hfill 193

\hspace{20pt}4.5.3 Li\'enard--Wiechert field in retarded 
coordinates\hfill 197

4.6 Radiation reaction via conservation laws\hfill 199

\hspace{20pt}4.6.1 Energy and momentum of the retarded 

\hspace{50pt}Li\'enard--Wiechert field\hfill 201

\hspace{20pt}4.6.2 Angular momentum of the retarded Li\'enard--Wiechert 
field\hfill 211

\hspace{20pt}4.6.3 Derivation of the Lorentz--Dirac equation\hfill 214

4.7 Conclusions\hfill 216

\vspace{10pt}

\noindent
{\bf 5. Radiation reaction in six dimensions}\hfill 223

5.1 Li\'enard--Wiechert potentials in 6D\hfill 225

5.2 Static potentials\hfill 228

\hspace{20pt}5.2.1 Conservation of electric charge in 6D\hfill 229

\hspace{20pt}5.2.2 {\it Static potential in various dimensions}\hfill 231

5.3 Retarded coordinates in six dimensions\hfill 232

5.4 Li\'enard--Wiechert field in 6D\hfill 233

5.5 Rigid point particle\hfill 234

5.6 Conservation laws\hfill 238

\hspace{20pt}5.6.1 Radiative part of energy-momentum\hfill 239

\hspace{20pt}5.6.2 Radiative part of angular momentum\hfill 241

5.7 Radiation reaction in 6D\hfill 242

5.8 {\it Bound parts of conserved quantities}\hfill 244

5.9 Conclusions\hfill 247

\vspace{10pt}

\noindent
{\bf 6. Interference in the radiation of two point-like charges}\hfill 251

6.1 Wheeler and Feynman electrodynamics\hfill 253

6.2 Two charged particles in classical field theory\hfill 262

6.3 Interference coordinate system\hfill 265

\hspace{20pt}6.3.1 Local map\hfill 267

\hspace{20pt}6.3.2 Global mapping\hfill 270

\hspace{20pt}6.3.3 Coordinate basis\hfill 274

6.4 Splitting of mixed part of the stress-energy tensor\hfill 277

6.5 Integration over time variables\hfill 282

6.6 Radiative part of interference energy-momentum\hfill 285

\hspace{20pt}6.6.1 Radiative part of interference angular momentum\hfill 291

6.7 Equations of motion of radiating charges\hfill 293

6.8 The problem of very close orbits\hfill 296

6.9 Conclusions\hfill 304

\vspace{10pt}

\noindent
{\bf 7. Self-force in $2+1$ electrodynamics}\hfill 321

7.1 Maxwell's equations in ${\mathbb M}_{\,3}$\hfill 322

\hspace{20pt}7.1.1 {\it Superfluid $^4$He film as 
$2+1$-electrodynamics}\hfill 328

7.2 Electromagnetic potentials in 2+1 theory\hfill 331

7.3 Electromagnetic field in 2+1 electrodynamics\hfill 333

\hspace{20pt}7.3.1 Static field\hfill 336

7.4 Violation of Huygens principle in $2+1$ electrodynamics\hfill 339

\hspace{20pt}7.4.1 Plane waves in ${\mathbb M}_{\,3}$\hfill 339

7.5 The flows of energy-momentum and angular momentum\hfill 342

\hspace{20pt}7.5.1 Coordinate system\hfill 346

\hspace{20pt}7.5.2 Energy-momentum tensor density\hfill 349

\hspace{20pt}7.5.3 New coordinate system\hfill 351

7.6 {\it Angular integration of energy-momentum tensor density}\hfill 356

7.7 {\it Integration over time variables}\hfill 360

7.8 Radiative parts of energy and momentum 

\hspace{20pt}carried by electromagnetic field\hfill 363

\hspace{20pt}7.8.1 Angular momentum of electromagnetic field\hfill 366

\hspace{20pt}7.8.2 Balance equations\hfill 366

\hspace{20pt}7.8.3 Renormalized mass of charged particle\hfill 368

7.9 Equation of motion of radiating charge\hfill 369

\hspace{20pt}7.9.1 Radiating charge in uniform static electric field\hfill 
371

7.10 Conclusions\hfill 373

\vspace{10pt}

\noindent
{\bf 8. Scalar self force}\hfill 377

8.1 Massless scalar field\hfill 378

\hspace{20pt}8.1.1 Energy-momentum and angular momentum\hfill 380

\hspace{20pt}8.1.2 Renormalization and radiation reaction\hfill 383

8.2 Green's function for a massive scalar field\hfill 385

\hspace{20pt}8.2.1 Scalar potentials\hfill 390

\hspace{20pt}8.2.2 Scalar field strengths\hfill 392

8.3 Scalar radiation\hfill 395

\hspace{20pt}8.3.1 Balance equations\hfill 397

\hspace{20pt}8.3.2 Renormalized mass of ``dressed'' scalar charge\hfill 399

\hspace{20pt}8.3.3 Effective equation of motion\hfill 401

8.4 Conclusions\hfill 402

\vspace{10pt}

\noindent
{\bf 9. Self-action problem for a massless charged particle}\hfill 405

9.1 Electromagnetic field of a photon-like charge\hfill 406

9.2 Energy-momentum and angular momentum\hfill 409

9.3 Ultrarelativistic limit of the Lorentz-Dirac equation\hfill 414

9.4 Conclusions\hfill 415

\vspace{10pt}

\noindent
{\bf References}\hfill 418

\chapter*{Introduction}

\markboth{Introduction}{Introduction}
\addcontentsline{toc}{chapter}{Introduction}\label{RR_Intro}

The aim of this book is to provide a self-contained and systematic 
introduction to problems of radiation and radiation reaction in classical 
field theory. The book is not intended to be a textbook in electrodynamics 
in the usual sense. We shall not attempt to exhibit neither a complete 
picture of the present state of classical particle electrodynamics nor the 
historical development of the subject to date. We focus it exclusively on 
the radiation phenomena in various models of classical field theory and 
on equations of motions of charged particles where the radiation reaction is 
taken into account.

Studies of the classical theory of charged particles and their radiation 
initiated by Lorentz and Abraham have demanded our attention over a 
century. (The early history of this subject is well presented in 
textbook \cite[Chapter 2]{Rohr}, see also Ref. \cite{Spohn}.) Abraham and 
Lorentz intended to develop a theory for the newly discovered electron.
The authors designed electron as a tiny charged sphere and summed up 
all the mutual electromagnetic forces acting between various charge 
elements. The total force was called the {\it self-force}. The first model 
of electron proposed by Abraham \cite{Abr1} was a rigid sphere with 
spherically symmetric charge distribution. (For a modern analysis see 
Ref. \cite{Spohn}.) The resulting expression for the self-force acting on 
Abraham's sphere with arbitrary charge's distribution was obtained by 
Lorentz \cite{Lrtz} in form of infinite series in powers of radius $R$ of 
the sphere. The first term is proportional to electron's acceleration and 
inversely proportional to its radius. The second term is proportional to the 
derivative of acceleration. It is structureless one while the higher terms 
all depend on the electron's structure, i.e. its radius $R$ and charge's 
distribution. Neglecting the higher order terms the author obtained 
generalization of the Newton's second law for a tiny charged sphere where the 
self-action is taken into account. The first term of the series which 
diverges as $1/R$ was coupled with particle's mass which then was proclaimed 
to be finite (the so-called {\it renormalization of mass}). Later Abraham 
\cite{Abrh} generalized the self-force expression to be valid for 
relativistic charged particle. Besides the proper time derivative of 
four-acceleration, his {\it radiation-reaction four-vector} contains also 
the Larmor term for a rate of energy-momentum radiated by oscillating 
charge. (Li\'enard and Heaviside generalized Larmor expression (1897) for a 
rate of radiation of non-relativistic charge on the case of relativistic 
one.)

Further progress in this field is associated with Dirac \cite{Dir}
who regarded the electron as an elementary particle with no internal 
structure. The author produced a proper relativistic derivation of the 
equation of motion of a point charged particle under the influence of an 
external force as well as its own electromagnetic field. Its commonly used 
and widely accepted title is the {\it Lorentz-Dirac equation}. To derive the equation Dirac \cite{Dir} used conservation 
of energy-momentum. This method is based on the conservation equation 
$div{\hat T}=0$ where ${\hat T}$ is the electromagnetic field 
energy-momentum tensor density. The differential statement of 
energy-momentum conservation can immediately be turned into integral 
statement. Applying Gauss's theorem the author calculate a flux of Maxwell 
energy-momentum tensor through a space-like surface enclosing a fragment of 
particle's world line. The work of an external force matches the flux of 
electromagnetic field energy-momentum and the change of particle's 
individual four-momentum. 

While the verification of energy-momentum conservation is not a trivial 
matter whenever we treat charges as point particles. Li\'enard-Wiechert 
fields are the solutions of wave equations with point-like sources. The 
fields as well as corresponding stress-energy tensor ${\hat T}$ get 
extremely large\footnote{\normalsize If a quantity tends to infinity when approaching 
particle's position, we call that it has a singularity at the position 
of the particle.} in the immediate vicinity of particle's world line. A 
real challenge is to reveal the part of ${\hat T}$ that produces finite 
flows of energy and momentum which detach themselves from the source and 
lead an independent existence. In his pioneer work \cite{Teit} Teitelboim 
splits the stress-energy tensor ${\hat T}$ of a point source into two parts 
which are separately conserved off the world line (see also review 
\cite{TVW}). Volume integration of the {\it radiative} part of the 
stress-energy tensor gives the integral of the Larmor relativistic rate of 
emitted energy-momentum over particle's world line. While the volume 
integration of the {\it bound} part results the term which is of essential 
value near the particle only. It describes the radiation that never goes 
far from the source and travels along with it. Angular momentum carried by electromagnetic field is also split into two parts with different properties: the divergent bound angular momentum depends on 
the instant characteristics of the charged particle while the emitted angular momentum accumulates with time \cite{LV}.

The bound part of the stress-energy tensor produces the ``cloud'' of energy 
and momentum which is permanently attached to the charge and is carried 
along with it. The bound part contains the self-interaction, which gives 
rise to the self-energy and self-stress of the source. It contributes to 
particle's inertia: four-momentum of charge contains, apart from usual 
velocity term, also a term which is proportional to the square of charge 
\cite[eq.(4.4)]{Teit}. A point-like singularity together with surrounding 
``cloud'' constitute new entity: {\it dressed charged particle}.

The Larmor relativistic rate of emitted energy-momentum describes the 
radiation which detaches itself from the source and leads an independent 
existence. The ``long-range'' radiation was the subject of active 
investigation since the early works of Schott and Schwinger \cite{Schw}. 
Schott describes the radiation emitted by circling electron \cite{Schot}. The 
radiation loses because of the high radial acceleration experienced by the 
electrically charged particles in circular accelerators set an upper limit 
to the attained energy. (Both cyclotron and synchrotron radiation have 
inherited their names from the devices used to accelerate charged particles 
in the 1930s and 1940s.) While radiation does more good than harm generally.
A sinuous beam of relativistic electrons which passes through a periodic 
magnetic field produced by arranging magnets with alternating poles is used 
as the amplification medium in {\it free electron lasers} \cite{FA}. The 
high intensity radiation emitted by this device can be tuned over a wide 
range of wavelengths. It is a very important advantage over conventional 
lasers.

The phenomenon of the self-force is intimately connected with the 
radiation; for this reason the ``self-force'' is also called 
``radiation-reaction force''. {\it The radiation removes energy, momentum, 
and angular momentum from the particle, which then undergoes a radiation 
reaction.} On the hypothesis that the total amounts of these quantities do not 
change we intend ``to find a formulation of classical 
charged particle theory which does not require any reference to, or 
assumptions about, the particle structure, its charge distribution, and its 
size.'' \cite[section 6.2]{Rohr}. 

The main feature of the present book is that it is 
exclusively based on symmetries and their associated conservation laws. 
Behavior of composite particles plus field system is governed by action 
principle which is invariant under infinitesimal transformations (rotations 
and translations) which constitute the Poincar\'e group. According to 
Noether's theorem, these symmetry properties yield conservation laws, i.e. 
these quantities that do not change with time. Conserved quantities place 
stringent requirements on the dynamics of the system. They demand that the 
change in field's energy, momentum, and angular momentum should be balanced 
by a corresponding change in the momentum and angular momentum of the 
particles, so that the total particles' plus field's momentum and angular 
momentum are properly conserved. The conservation laws are an immovable 
fulcrum about which tips the balance of truth regarding renormalization and 
radiation reaction.

We now give an outline of this book. 

Its structure is logical and didactic rather than historical. The necessary 
mathematical tools are presented in the first two chapters. Chapter 1
contains elementary differential geometry. It is oriented toward physicists 
who are not expert in this field. This Chapter covers manifolds and 
their coordinatizations, vector fields and tangent bundles, Cartan's 
differential forms and exterior differentiation, Lie groups and Lie 
algebras etc. A special attention is paid to Stoke's theorem and 
integration. Chapter 2 contains some aspects of classical electrodynamics. 
Section 2.4 presents Maxwell's equations in coordinate-free 
form. The equations look the same not only in any coordinates but also in 
Minkowski space of arbitrary dimensions where the sources and the field 
``live''. Sections 2.5 and 2.6 review the Lagrangian 
formalism of classical field theory, Noether theorem and conservation laws. 
In Section 2.7 the algebraic properties of the electromagnetic 
field tensor are analyzed.
 
Chapter 3 covers solutions to Maxwell's equation in Minkowski space of 
various dimensions. We use the Green's function technique to solve the 
inhomogeneous wave equation, i.e. equation of motion for the electromagnetic 
field. In general, a Green's function consists of ``direct'' and ``tail'' 
parts which are proportional to the Dirac's delta-function and to the Heaviside 
stepfunction, respectively. The ``direct'' part results the electromagnetic field at a 
point $x$ which is determined completely by state of motion of a source on 
the past null cone with vertex at $x$. It is true for the 
Li\'enard-Wiechert potentials in Minkowski space of 
even dimensions, including four-dimensional spacetime. In odd dimensions 
the ``tail'' part of Green's function arises which allows for additional 
contribution from source {\it within} the past null cone of $x$ where the field 
is evaluated.

Chapter 4 treats self-interaction in four-dimensional electrodynamics. The 
Li\'enard-Wiechert solutions to wave equations (both the retarded and the 
advanced) are given in the first two sections. Much attention is paid to 
their symmetry properties with respect to time inversion. In Section 4.4 we 
present Dirac's regularization procedure based on the decomposition of the 
retarded Li\'enard-Wiechert field into the singular ``mean of the advanced 
and retarded field'' and the finite ``radiation'' field that reaches a 
distant sphere. While the main feature of this book is that we prefer the 
retarded fields as those of true physical meaning. Following our 
experience, we suppose that point charged particles carry with them fields 
that behave as outgoing radiation. In Section 4.5 we describe 
the retarded coordinate systems which are used  for calculations of flows 
of electromagnetic field energy-momentum and angular momentum. We compare 
Dirac's retarded coordinates \cite{Dir} and Bhabha's coordinates \cite{Bha} 
associated with non-inertial reference frame that travels along with an 
accelerated charge. We find the energy-momentum and angular momentum 
emitted by the charge as well as their bound counterparts which are not 
emitted but remain linked to the charge. Much of this Chapter is drawn from the research literature and some of it 
appears to be new. Particularly noteworthy in Section 4.6 which presents a
derivation of the Lorentz-Dirac equation which relies on ten conserved 
quantities corresponding to Poincar\'e-invariance of a composite particle 
plus field system. 

Chapter 5 explores further application of the renormalization procedure 
based on energy-momentum and angular momentum balance equations. It deals 
with the self-action problem for a point-like charge arbitrarily moving in 
flat spacetime of six dimensions. Electromagnetic field tensor consists of 
electric field with five components and magnetic field with ten components. 
In Section 5.1 we present the solution to wave equation with 
point-like source. Six-dimensional analog of the Li\'enard-Wiechert 
potential depends not only on particle's position and six-velocity, but 
also on its six-acceleration. Coulomb electrostatic potential scales as 
$|{\mathbf r}|^{-3}$ in six dimensions (see Section 5.2 where 
static potentials in arbitrary dimensional spacetime are given). Inevitable 
infinities arising in six-dimensional electrodynamics are stronger than in 
four dimensions and cannot be removed by the renormalization of mass. A 
closer look on balance equations explores that the ``bare'' singularity 
possesses something like internal angular momentum. Its magnitude is 
proportional to the norm of six-acceleration. It is the so-called {\it 
rigid} relativistic particle described in Section 5.5. Its 
motion is governed by the higher derivative Lagrangian depending not only 
on the arclength of the world line, but also on its curvature. Surrounding 
electromagnetic ``cloud'' contributes in six-momentum of dressed charge, so 
that dynamics of the dressed charged particle is reacher that of ``bare'' 
singularity. In Section 5.7 we derive six-dimensional analog of 
the Lorentz-Dirac equation which was firstly obtained by Kosyakov 
\cite{Kos}. Analysis of balance equations is the cornerstone of our 
treatment of the regularization procedure.

Chapter 6 discusses two-body problem in conventional electrodynamics where 
interference in the radiation of two point-like charges is taken into 
account. We determine ten conserved quantities corresponding to Poincar\'e 
symmetry of a closed system of two charges and their electromagnetic field.
Since the stress-energy tensor is quadratic in field strengths and the field 
satisfies superposition principle, the tensor contains mixed part which 
describes interference of outgoing electromagnetic waves from arbitrarily 
moving point-like sources. The so-called {\it direct particle fields} 
\cite{HN} arise due to volume integration of mixed part of two-particle 
stress-energy tensor. These direct fields (retarded and advanced) are 
functionals of particles' world lines; they do not possess degrees of 
freedom of their own. Therefore, we arrive at the realm of {\it 
action-at-a-distance} electrodynamics \cite{HN}.

The theory was elaborated by Wheeler and Feynman \cite{WF1,WF2}. It is based
on the following assumptions \cite[p.160]{WF1}:
\begin{itemize}
\item
An accelerated point charge in otherwise charge-free space does not radiate
electromagnetic energy.
\item
The fields which act on a given particle arise only from other particles.
\item
These fields are represented by one-half the retarded plus one-half the
advanced Li\'enard-Wiechert solutions of Maxwell's equations. This law is
symmetric with respect to past and future.
\item
Sufficiently many particles are present to absorb completely the radiation
given off by the source.
\end{itemize}

Since the source emanates in all possible directions, all the particles of
the universe are required to absorb completely the radiation. They
constitute a {\it perfect absorber} which possesses a remarkable twofold
property: it cancels the (acausal) advanced part of the fields acted on a
given particle and doubles the retarded one. Therefore, the complete
absorption is the crucial issue of the theory. For this reason Wheeler and
Feynman called it the {\it absorber theory of radiation}.

Rigorous calculations performed in Chapter 6 reveal that the
combination of retarded Li\'enard-Wiechert fields forms a resultant field
with the desired properties. Then the ``perfect absorption'' should be
replaced by the interference of outgoing waves in Wheeler and Feynman
electrodynamics. It allows us to reconcile Wheeler and Feynman theory with
the concept of retarded causality.

The Lorentz-Dirac equation governs the motion of a point-like charge in 
flat spacetime. DeWitt and Brehme \cite{WB} generalized Dirac's work to 
a curved spacetime. (Their expression for the self-force acting on a point 
charge radiating electromagnetic waves in a curved space background was 
later corrected by Hobbs \cite{Hb}.) Background gravitational field ``slows 
down'' photons mediating electromagnetic interaction, so that they move 
with all velocities smaller than or equal to the speed of light. For this 
reason the charge ``fills'' its own field, which acts on it just like an 
external one. In the present book we do not consider electrodynamics in 
curved spacetime. Instead we analyse in Chapter 7 the self-action problem 
for an electric charge arbitrarily moving in flat spacetime of three dimensions. An 
essential feature of $2+1$ electrodynamics is that the radiation develops a 
tail, as it is in four-dimensional curved spacetime. This is because 
the retarded Green's function associated with D'Alembert operator is 
supported within the light cone (see Chapter 3, paragraph 3.4.1). 

In Section 7.1 we present the Maxwell's equations in Minkowski 
space of three dimensions. Electromagnetic field tensor consists of 
electric field with two components and scalar magnetic field. We outline 
the remarkable correspondence between the Maxwell's equations and equations 
that govern behaviour of superfluid $\!\!\!\phantom{I}^4{\rm He}$ film. By 
this we mean that the dynamics of the low energy quasiparticles and 
elementary excitations living inside a helium film is governed by Maxwell's 
equations in ${\Bbb M}_3$. Therefore, if one study the behavior of electric 
charges living inside hypothetic spacetime with two space directions, they 
study the kinetic of vortices and phonon excitations in superfluid 
$\phantom{}^4{\rm He}$ film.

Computation of electromagnetic field energy, momentum, and angular momentum 
carried by the tail field is highly non-trivial because outgoing waves 
emitted by different points of particle's world line combine with one 
another. This phenomenon is called the violation of Huygens principle; it 
is described in Section 7.4. In Section 7.5 we introduce a 
coordinate system which allows to calculate in Sections 7.6, 7.7, and 
7.8 the total flows of (retarded) Noether quantities which 
flow across the plane $x^0=t$ with fixed $t$. From the analysis of balance 
equations we develop the regularization procedure for a theory where 
Green's function involves a ``tail'' part. The renormalization scheme 
manipulates fields on the world line only. It generalizes Dirac's scheme of 
decomposition of the direct retarded field into the singular ``mean of the 
advanced and retarded field'' and the finite ``radiation'' field. In 
Section 7.9 we derive integro-differential equation which plays 
the role the Lorentz-Dirac equation in three dimensions. The radiation 
reaction is determined by the Lorentz force of point-like charge acting 
upon itself plus a term that
provides finitness of the self-action. The word ``self-force'' can be 
understood literally: the charge in three dimensions is repulsed by itself 
taken in the past. The integro-differential equation of motion contains path 
integral over the past motion of the dressed charge.

Chapter 8 explores application of the renormalization procedure for 
theories with tail fields to a point-like source coupled with neutral 
massive scalar field. Corresponding equations of motion were first found by 
Bhabha \cite{Bha} following a method originally developed by Dirac 
\cite{Dir} for the case of electromagnetic field. In this method the finite 
force and self-force terms in the equations of motion are obtained from the 
conservation laws for the energy-momentum tensor of the field. It was 
extended by Bhabha and Harish-Chandra \cite{BHC} to particles interacting 
with any tensor field and was applied to the motion of a simple 
pole of massive scalar field by Harish-Chandra \cite{HC}. 

Recently \cite{Q}, Quinn has obtained an expression for the self-force on a 
point-like particle coupled to a massless scalar field arbitrarily moving in 
a curved spacetime. Quinn establishes that the total work done by the scalar 
self-force matches the amount of energy radiated away by the particle. To 
cancel a troublesome part of self-force near the upper limit of path 
integral, the author averages diverging piece over a small, spatial 
two-sphere surrounding the particle.

In Chapter 8, we split the energy-momentum and angular momentum carried 
by massive scalar field into bound and radiative parts. Extracting of 
radiated portions of Noether quantities is not a trivial matter, since the 
massive field holds energy and momentum near the source. We apply the 
procedure elaborated in preceding Chapter, where $2+1$ electrodynamics has 
been considered. To substantiate the splitting, we analyze asymptotic 
behavior of the bound and the radiative pieces as well as the total 
energy-momentum and angular momentum balance equations. It is of great 
importance that conservation laws yield the Harish-Chandra equation of 
motion. This circumstance reinforces our conviction that Dirac's 
decomposition scheme can be applied to tail fields. We regard this approach 
as preferable, because radiative terms are smooth at the location of the 
particle, so that averaging is not required in computing the self-force.  

In Chapters 4-8 the rearangement of the initial degrees of 
freedom in classical field theory has been perforned which leads to new 
dynamical entities: dressed charged particles and radiation. A dressed 
charged particle is a ``bare'' source surrounded by neighbouring field's 
``cloud'' while the radiation is the ``far'' field which detaches the 
source and leads to independent existence. All the problems considered 
before are renormalizable: unavoidable divergences are absorbed within the 
renormalization procedure. The renormalization modifies particles' 
individual characteristics. In Chapter 9 we consider an 
interesting example of non-renormalizable theory: classical electrodynamics 
of massless charged particles \cite{Bonnor,Dolan}. Unlike the massive case, 
the charge having zero rest mass is not ``dressed''. By this we mean that 
the photon-like charge does not possess any electromagnetic ``cloud'' 
permanently attached to it. It produces the far field which diverges not 
only on particle's world line, but also at points on light ray that extends 
to infinity. The ``ray singularity'' can not be removed by means of 
renormalization of particle's individual characteristics.

Over a sixty figures, tables and diagrams illustrate the book. We hope that it helps to explain the meaning of material, because there are many people for whom geometrical reasoning is easier than purely analytic reasoning. To provide guidance to the book's contents, section headings within chapters are pointed in two different styles. Fundamental material is marked by {\bf boldface} headings,
while supplementary topics are marked by {\it italics}.

\end{document}